\documentclass[twocolumn]{aastex62}
\usepackage{gensymb}
\shorttitle{Electron irradiation of CO ices}
\usepackage[version=4]{mhchem}
\usepackage{float}
\usepackage{verbatim}
\usepackage{amsmath}
\usepackage{soul}
\usepackage{hyperref}

\graphicspath{{./}{figures/}}

\received{January 1, 2018}
\revised{January 7, 2018}
\accepted{\today}
\submitjournal{ApJ}

\shortauthors{C.-H. Huang et al.}

\begin{document}
\title{Effects of $150-1000$~eV Electron Impacts on Pure Carbon Monoxide Ices using \\ the Interstellar Energetic-Process System (IEPS)}

\author[0000-0003-2741-2833]{C.-H. Huang}
\affiliation{Department of Physics, National Central University, Jhongli City, Taoyuan County 32054, Taiwan;\\
\href{mailto:chhuang@phy.ncu.edu.tw}{chhuang@phy.ncu.edu.tw},\href{mailto:asperchen@phy.ncu.edu.tw}{$^{\star}$asperchen@phy.ncu.edu.tw}}
\author{A. Ciaravella}
\affiliation{INAF - Osservatorio Astronomico di Palermo, P.za Parlamento 1, 90134 Palermo, Italy}
\author{C. Cecchi-Pestellini}
\affiliation{INAF - Osservatorio Astronomico di Palermo, P.za Parlamento 1, 90134 Palermo, Italy}
\author{A. Jim\'enez-Escobar}
\affiliation{INAF - Osservatorio Astronomico di Palermo, P.za Parlamento 1, 90134 Palermo, Italy}
\author{L.-C. Hsiao}
\affiliation{Department of Physics, National Central University, Jhongli City, Taoyuan County 32054, Taiwan;\\
\href{mailto:chhuang@phy.ncu.edu.tw}{chhuang@phy.ncu.edu.tw},\href{mailto:asperchen@phy.ncu.edu.tw}{$^{\star}$asperchen@phy.ncu.edu.tw}}
\author{C.-C. Huang}
\affiliation{Department of Physics, National Central University, Jhongli City, Taoyuan County 32054, Taiwan;\\
\href{mailto:chhuang@phy.ncu.edu.tw}{chhuang@phy.ncu.edu.tw},\href{mailto:asperchen@phy.ncu.edu.tw}{$^{\star}$asperchen@phy.ncu.edu.tw}}
\author{P.-C. Chen}
\affiliation{Department of Physics, National Central University, Jhongli City, Taoyuan County 32054, Taiwan;\\
\href{mailto:chhuang@phy.ncu.edu.tw}{chhuang@phy.ncu.edu.tw},\href{mailto:asperchen@phy.ncu.edu.tw}{$^{\star}$asperchen@phy.ncu.edu.tw}}
\author{N.-E. Sie}
\affiliation{Department of Physics, National Central University, Jhongli City, Taoyuan County 32054, Taiwan;\\
\href{mailto:chhuang@phy.ncu.edu.tw}{chhuang@phy.ncu.edu.tw},\href{mailto:asperchen@phy.ncu.edu.tw}{$^{\star}$asperchen@phy.ncu.edu.tw}}
\author{Y.-J. Chen$^{\star,}$}
\affiliation{Department of Physics, National Central University, Jhongli City, Taoyuan County 32054, Taiwan;\\
\href{mailto:chhuang@phy.ncu.edu.tw}{chhuang@phy.ncu.edu.tw},\href{mailto:asperchen@phy.ncu.edu.tw}{$^{\star}$asperchen@phy.ncu.edu.tw}}

\begin{abstract}
Pure CO ice has been irradiated with electrons of energy in the range $150-1000$~eV with the Interstellar Energetic-Process System (IEPS). The main products of irradiation are carbon chains C$_n$ ($n=3$, 5, 6, 8, 9, 10, 11, 12), suboxides, C$_n$O ($n=2$, 3, 4, 5, 6, 7), and C$_n$O$_2$ ($n=1$, 3, 4, 5, 7) species. \ce{CO2} is by far the most abundant reaction product in all the experiments. The destruction cross-section of CO peaks at about 250 eV, decreases with the energy of the electrons and is more than one order of magnitude higher than for gas-phase CO ionization. The production cross-section of carbon dioxide has been also derived and is characterized by the competition between chemistry and desorption. 

Desorption of CO and of new species during the radiolysis follows the electron distribution in the ice. Low energy electrons having short penetration depths induce significant desorption. Finally, as the ice thickness approaches the electron penetration depth the abundance of the products starts to saturate. Implications on the atmospheric photochemistry of cold planets hosting surface CO ices are also discussed.  
\end{abstract}

\keywords{astrochemistry  --- protoplanetary disks --- methods: laboratory --    molecular processes}

\section{Introduction} \label{intr}
Carbon monoxide is the dominant carbon reservoir in molecular gas, and among the most abundant molecules in the Universe. Interstellar CO ice, as distinct from gaseous molecules, was first positively identified in 1984 with the detection of infrared absorption at 4.67 $\mu$m in several molecular clouds \citep{Lacy84}. Such detection occurred a decade after the discovery  of interstellar water ice observed towards the embedded protostellar Becklin-Neugebauer object by \citet{GF73}. Subsequent observations suggested that ices containing CO were ubiquitous in cold interstellar and circumstellar regions \citep{HvD09}. These astronomical observations stimulated intense activity in laboratory studies of solid-state chemistry, that together with accurate observational results gave rise to the current description of ices in space. We know that the onset of CO ice occurs in darker regions than those in which the onset of water ice occurs \citep{Boogert15}. Water ice deposition is mostly completed before the CO ice is deposited on the dust grains. Therefore, interstellar ices appear to have a layered structure surrounding the dust grain cores:  an inner water-rich layer forms early in the evolution of a cloud by hydrogenation of oxygen, frozen at the grain surface, and solid methane and ammonia are also likely to be formed in surface reactions at this stage, through hydrogenation of carbon and nitrogen. During the initial stages of star formation, the increase in gas density is so steep that the freeze-out timescale shortens dramatically, inducing a catastrophic removal of the gas, whose main component is carbon monoxide. The formation of the CO ice layer provides the raw material from which icy methanol and other species may be formed \citep{Ch16}. 

Carbon monoxide is known to be present as ice on planetary and moon surfaces such as Triton \citep{L10}  ~\textendash~ presumably a captured Kuiper Belt object ~\textendash~ and Pluto \citep{BF16}, and supposed to be present in exoplanetary cryospheres \citep{Be06}. 

Many types of radiation fields ubiquitous in interstellar and circumstellar regions can chemically process carbon monoxide giving rise to a variety of effects going from chemical evolution to photodesorption. \citet{Jam06} present a detailed discussion of energetic particle bombardment. To simulate the energetic electrons trapped in magnetospheres of planets and to reproduce the irradiation effects of typical Galactic cosmic ray particles, these authors performed experiments in which CO ices kept at cryogenic temperatures were irradiated with (moderately) energetic keV electrons. In this work, we exploit electrons in the energy middle-range ($150-1000$~eV), to explore the effects at energies closer to mean energies of primary electrons produced by cosmic-rays \citep{CCP92} and X-rays \citep{T97} impacting with molecular gas.

The electrons used in this work lose 99\% of their energy within $52.5$~nm inside the CO ice. This penetration depth is much shorter than the 690~nm required for the 5~keV electrons employed by \citet{Jam06}, the $\sim 440$~nm of the ultraviolet photons (\citealt{CD}), and the $\sim 2500$~nm of 550~eV X-rays \citep{Cia16}.

In Section~\ref{IEPS} we present the experimental facility used in this work. We describe our experiments in Section~\ref{CO_Exp}, illustrate the results in Section~\ref{analysis}, and discuss them in Section~\ref{disc}. The last Section contains our conclusions.

\section{Experimental Facility} \label{IEPS}
\begin{figure*}
\centering
\includegraphics[width=14.5cm]{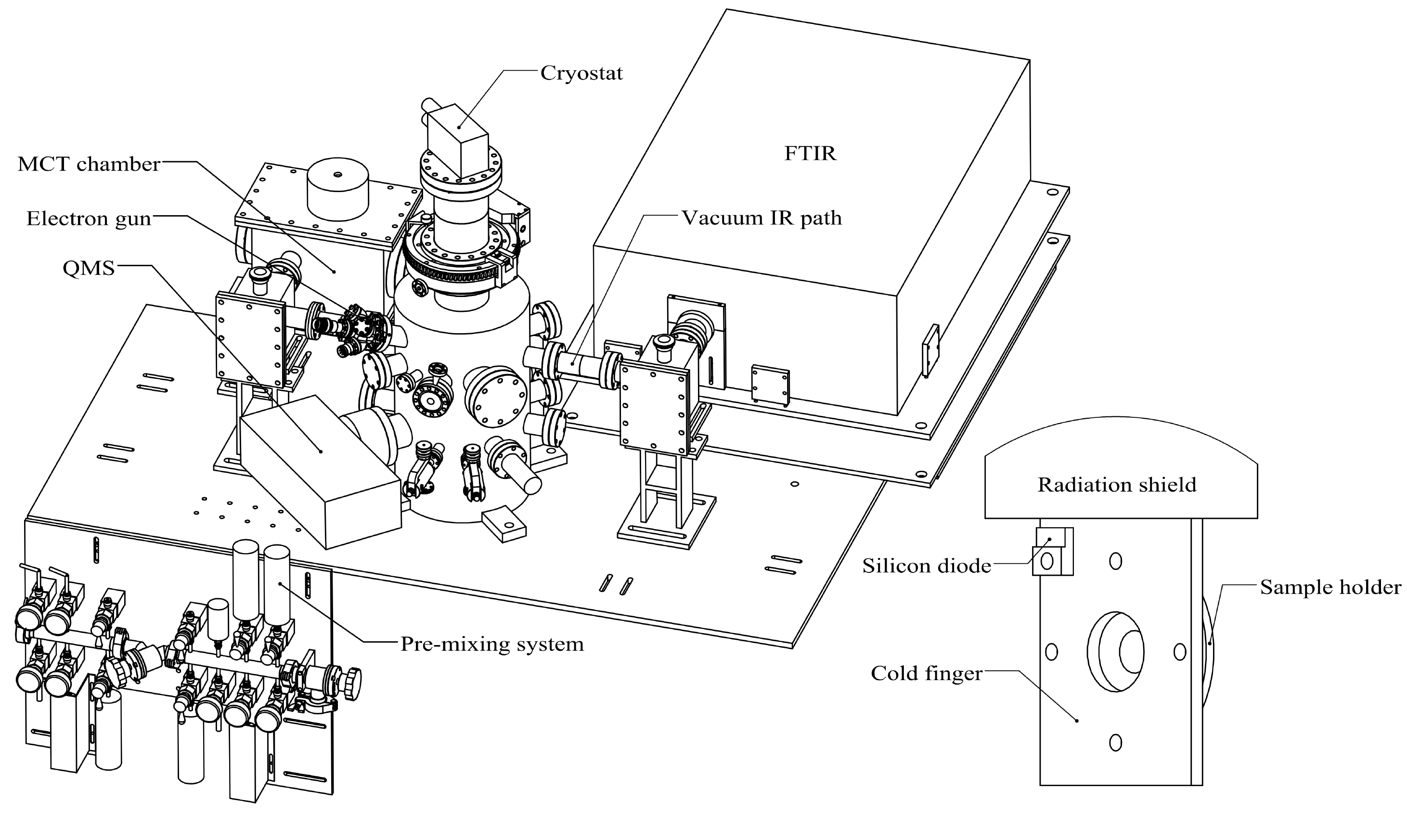}
\caption{IEPS UHV experimental setup. In the right inset some details of the cold finger assembly are reported.}
\label{IEPS_illu}
\end{figure*}

The experiments reported in this work have been performed with the Interstellar Energetic-Process System (IEPS, National Central University, Taiwan). IEPS is an ultra-high vacuum (UHV) chamber equipped with radiative sources, a pre-mixing system, and detection/diagnostic instruments. A schematic picture of IEPS is shown in Figure~\ref{IEPS_illu}. The facility is detailed below.

\subsection{Ultra High Vacuum Chamber} \label{UHV}
The base pressure inside IEPS, measured by an ion gauge (Granville-Phillips 370 Stabil), is $\lesssim 5 \times 10^{-10}$ torr. The vacuum is obtained with a turbo-molecular pump (Oerlikon Leybold vacuum 600C, 600 l~s$^{-1}$),  backed up by a scroll pump (Edward nXDS15i). A closed-cycle helium cryostat (CTI M350), is mounted on a rotating platform (Thermionics RNN-400 equipped with stepper motor) at the center of the chamber, see Figure~\ref{IEPS_illu}. At the end of the cold finger there is a sample holder (see the inset in Figure~\ref{IEPS_illu}), where an infrared transparent window can be mounted as substrate for the ices. Two temperature sensors driven by a Lakeshore 335 temperature controller are mounted along the cold finger. The first one is close to the sample holder (Lakeshore DT-670-OB), while the other is located on tip of the second stage of the cryostat (Lakeshore DT-470-SD). The cold finger and copper radiation shield are placed on the second and first stages of the cryostat, respectively. A heater is hooped on the cold finger and operated through the temperature controller. In order to improve thermal conductivity, all junctions are made with Indium (99.99\% purity). The infrared transparent window is housed in the sample holder of the cryostat. The window is held by a circular copper ring tightened by screws with spring washers to avoid damaging the window at low temperatures. In this configuration, the substrate can be maintained at temperatures in the range of $11 - 340$~K with an uncertainty of $\pm 0.3$~K. The infrared and ultra-violet sources are $45\degree$ incident on the sample holder. IEPS is designed to allow simultaneous irradiation with different energetic sources. In the present configuration, there are an electron gun (Kimball physics EFG-7 equipped with EGPS-2017 power supply) and a T-type \citep{CH14} microwave-discharge hydrogen-flow lamp (Opthos Instruments, INC.).

\subsection{Pre-mixing System} \label{Pre-mix}
Gas mixtures are prepared in the  stainless steel pre-mixing system made of two  gas-lines containing four bottles of equal volumes. The system is pumped by a turbo-molecular pump (Oerlikon Leybold 361, 400 l~s$^{-1}$) backed up by an oil-sealed mechanical pump (Alcatel 2012A) equipped with an oil trap containing molecular sieves (type 13X). The system is baked at $100\celsius$~ to obtain optimal vacuum conditions, and eliminate contamination. The base pressure is  $8 \times 10^{-8}$~torr. The partial pressures of the gas components are measured by a capacitance manometer (BROOKS CMC series working in the range of $0-100$~torr with 0.5\% accuracy). The prepared gas sample is introduced into the UHV chamber through a leak valve (VG, ZLVM 940R) connected to a 1/16-inch stainless tube pointed toward the cold substrate.

\subsection{Detection System} \label{Detector}
A Fourier Transform (mid-)InfraRed (FTIR) spectrometer (Bruker VERTEX 70), and a Quadrupole Mass Spectrometer (QMS, Hiden analysis/3F RC PIC) are the diagnostic instruments. The FTIR spectrometer is equipped with a Mercury-Cadmium-Tellurium (MCT) detector records trasmission spectra of the ice sample. To reduce the absorbance of gas phase CO$_2$ and H$_2$O in the atmosphere, the infrared beam is introduced into a vacuum path separated from UHV by two wedged ZnSe window flanges, located at the entrance and exit sides of the UHV chamber. The MCT detector is kept in a rectangular vacuum chamber. The QMS is mounted with its central axis pointed along the normal direction of the substrate. The emission current of the QMS filament and integrated scanning time can be manually regulated. The exploited emission current is $100~\mu$A, and the scanning time $100$~ms for each mass. The electron-impact ionization energy has been chosen to be 70~eV. The minimal partial pressure detectable by the QMS is $3.5 \times 10^{-15}$~torr. 

\section{Experiments} \label{CO_Exp}
The experiments consist of three phases: deposition, irradiation, and warm-up. The CO gas (CINGFONG GAS INDUSTRIAL, purity 99.99\%), prepared in the pre-mixing chamber at 5 torr, is introduced in the UHV chamber at a pressure of $10^{-7}$ torr for about 20 minutes, and condensed on a CaF$_2$ window cooled at 11 K. At the end of the deposition we wait 30 min in order to obtain vacuum conditions similar to those before deposition.

We compute the ice thickness $\Delta L$ using the 2136 cm$^{-1}$ feature of CO with a band strength $A = 1.1 \times 10^{-17}$~cm~molecules$^{-1}$ \citep{Jia75}. The CO column density measured in monolayers $(1~{\rm ML} = 1 \times 10^{15}$~molecules cm$^{-2})$ along the infrared beam is calculated as
\begin{equation}
\Delta L_{\rm IR} = \frac{1}{A}  \int_{\Delta \nu} \tau_\nu d\nu
\end{equation}
where $\tau_\nu$ is the optical depth of the band, and $\nu$~cm$^{-1}$ the wavenumber. In all the experiments the ice thickness is measured to be $\Delta L = 707$ ($319$~nm). The infrared and electron beams impinge the ice surface at $45\degree$ and $25\degree$, respectively. Therefore, the effective ice thickness  is  $1000$~ML ($452$~nm) along the infrared beam, and $781$~ML ($353$~nm) along the electrons beam. In the following, unless otherwise specified, column densities refer to the direction along the electron beam. The major error on the ice thickness is dominated by the uncertainty in the band strength, i.e. 9.5\% \citep{Jia75}.


The infrared spectra collected before and during the irradiation are obtained using 2~cm$^{-1}$ resolution. At the end of the radiolysis the ice is  warmed up to room temperature at a rate of 2~K~min$^{-1}$. During this phase, infrared spectra have been collected every 5~K starting from 15~K, with 2~cm$^{-1}$ resolution. 

The QMS monitored gas species in the chamber from $m/z = 12$ to 100, for the entire duration of the experiments. During the warm-up the mass $m/z = 28$, related to CO, has been excluded because its intensity can be higher than the safe ion counts limit ($10^7$ ion counts~s$^{-1}$) of the QMS.

The radiolysis of pure CO ice is performed using electrons of energies from $E_e = 150$ to 1000 eV, and a flux of $f = 3.2 \times 10^{11}$~electron~cm$^{-2}$~s$^{-1}$. The energy uncertainty is 0.02\%. The spot size is 2~cm$^2$. The irradiation time is chosen to have similar total impinging energy in all the experiments.  For the lowest energy case, $E_e =150$~eV we stopped the irradiation earlier because the abundances of products do not vary appreciably after three long ($\Delta t = 1000$~s) irradiation steps.

The experiments, repeated at least two times, are summarized in Table~\ref{exp}. The columns in the table list the electron energies used in the experiments, the total irradiation times, the total absorbed electrons, the total absorbed energies, and the penetration length along the direction of the electron beam, $\Lambda_e$. We characterize each electron energy through a penetration length defined as the depth within the ice at which the electrons loose 99\% of their initial energies. The penetration lengths are estimated with "monte CArlo SImulation of electroN trajectory in sOlids" (CASINO) software \citep{Dr07} using a CO ice density $\delta_{\rm CO} = 1.0288$~g~cm$^{-3}$ \citep{Jia75}. The statistic error on the penetration depth is 0.1\%.

\begin{deluxetable*}{cccccc}
\tabletypesize{\scriptsize} 
\tablecaption{Experimental parameters\label{exp}}
\tablewidth{40pt}
\tablehead{\\
 Electr. Energy & Irrad. Time & Dose & Abs. Elec. & Abs. Ener. & $\Lambda_e$ \\
 (eV) & (s) & (eV molecule$^{-1}$) & (elec cm$^{-2}$) & (eV cm$^{-2})$ & (ML/nm) }
\startdata
150 &3095 &27.5 &$1.0 \times 10^{15}$ &$1.5 \times 10^{17}$ &8.5/3.8\\
200 &6095 &45.6 &$1.9 \times 10^{15}$ &$3.9 \times 10^{17}$ &12.8/5.8\\
250 &4715 &32.8 &$1.5 \times 10^{15}$ &$3.8 \times 10^{17}$ &17.2/7.8\\
300 &3095 &20.4 &$1.1 \times 10^{15}$ &$3.3 \times 10^{17}$ &21.8/9.9\\
400 &2675 &16.2 &$8.7 \times 10^{14}$ &$3.5 \times 10^{17}$ &31.8/14.4\\
500 &2305 &13.0 &$7.6 \times 10^{14}$ &$3.8 \times 10^{17}$ &42.8/19.3\\
600 &1985 &10.3 &$6.5 \times 10^{14}$ &$3.9 \times 10^{17}$ &55.3/25.0\\
800 &1465 &6.7  &$4.8 \times 10^{14}$ &$3.9 \times 10^{17}$ &83.6/37.8\\
1000 &1255 &5.2 &$4.1 \times 10^{14}$ &$4.1 \times 10^{17}$ &116.2/52.5\\
\enddata
\end{deluxetable*} 

\section{Results and Analysis}\label{analysis}
\subsection{Products of the Irradiations}
\begin{figure*}
\centering
\includegraphics[width=14.cm]{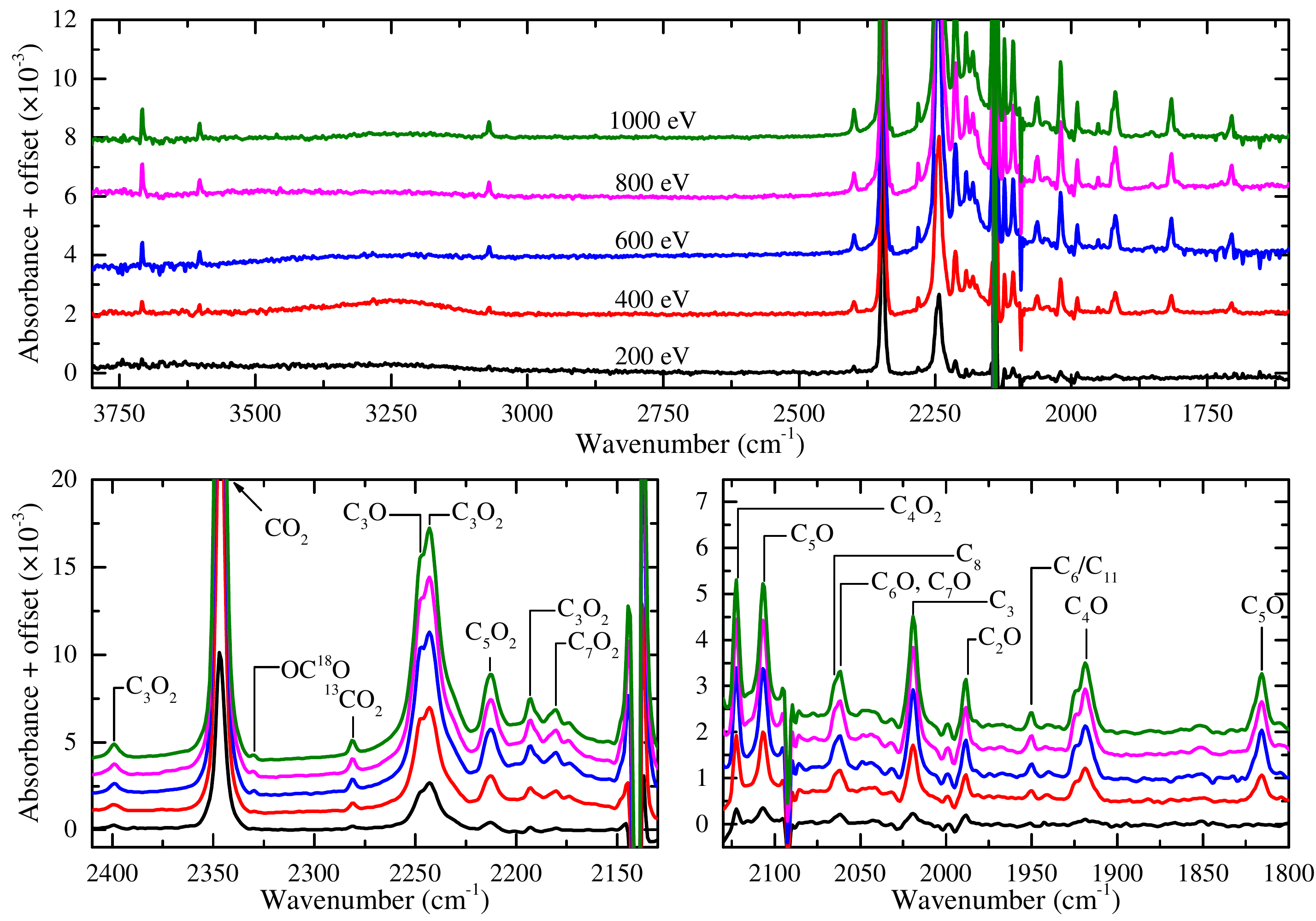}
\caption{Infrared spectra of the CO ice after radiolysis with electrons of 200 (black), 400 (red), 600 (blue), 800 (pink), and 1000 (dark green) eV. The spectra are obtained as differences between the irradiated and the  sample spectra. The top panel shows the spectra in the range of 3800 to 1600~cm$^{-1}$, while the bottom panels show the detailed region from 2410 to 1800~cm$^{-1}$. Main products are marked.}
\label{IR_Spc}
\end{figure*}

Figure~\ref{IR_Spc} shows the infrared spectra after the CO ices has been irradiated with the same number of impinging electrons, $2.2 \times 10^{14}$~electrons~cm$^{-2}$. 

The impinging electrons ionize CO molecules, producing a cascade of secondary electrons. The net effect is ionization, excitation and dissociation of CO molecules, and the injection of reactive ions, atoms and molecules into the ice. Below 10~eV the CO molecules is destroyed by dissociative electron attachment \citep{LP16}. The products of irradiation can be sorted in three groups: carbon chains C$_n$, suboxides, C$_n$O, and C$_n$O$_2$ species (details on the reaction scheme can be found in \citealt{Jam06}). \ce{CO2} ($\nu_3$ at 2346 cm$^{-1}$) is the most abundant product in all the experiments. The features at 2248, 2243, and 1981 cm$^{-1}$, are assigned to \ce{C3O}, \ce{C3O2}, and \ce{C2O}, respectively. They are common to ultraviolet irradiation of a thick CO ice \citep{Ge96}, X-ray irradiation \citep{Cia16}, and 5 keV radiolysis \citep{Jam06}. In this latter experiment, longer chain products such as \ce{C5O2} and \ce{C7O} are also observed. The detection of \ce{C7O2} at 2115 cm$^{-1}$ has been also reported in soft X-ray irradiation experiments of pure CO ices \citep{Cia16}. The non-detection of \ce{C7O2} by \citet{Jam06} may be due to the blending of this feature with the CO-Ag band at 2112~cm$^{-1}$. The carbon chains \ce{C8} and \ce{C9} are obtained in X-ray irradiation experiments, but not mentioned in \citet{Jam06}. Finally the unknown features at 1578, 2045, 2049, 2153, 2173, and 2235 cm$^{-1}$ have been also reported in \citet{Cia16}. The products of the irradiation are summarized in Table~\ref{Assi}, where we report the wavenumber, the assignment,  the band strength (whenever available), the lowest energy experiments in which the product has been  detected, and the references.

\begin{figure*}
\centering
\includegraphics[width=16cm]{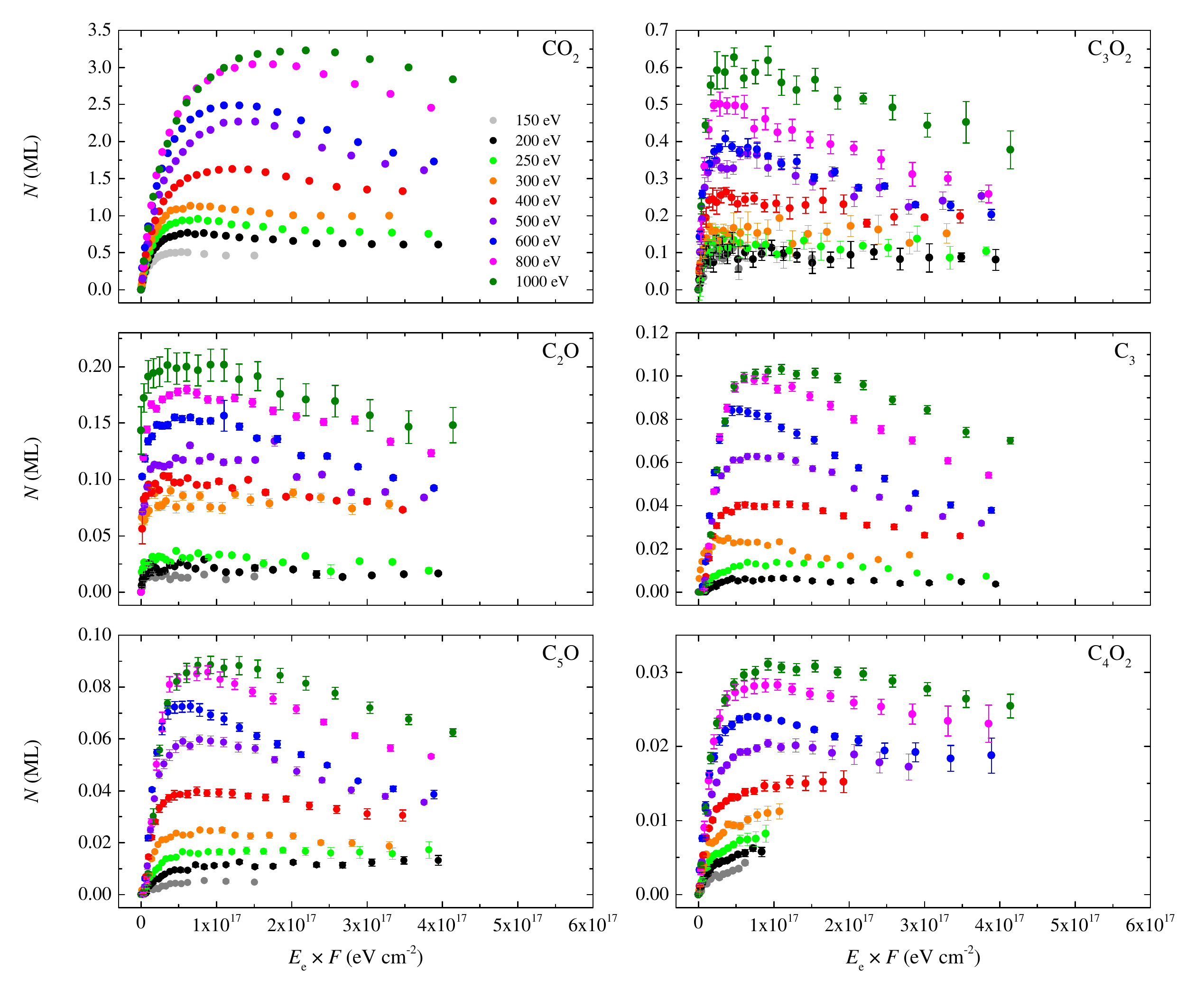}
\caption{Column density versus absorbed energy for \ce{CO2}, \ce{C3O2}, \ce{C2O}, \ce{C3}, \ce{C5O} and \ce{C4O2}.  The different colors are for electron of 150 (grey dots), 200 (black), 250 (light green), 300 (orange), 400 (red), 500 (purple), 600 (blue), 800 (pink), and 1000 eV (dark green).}
\label{Fig_prods}
\end{figure*}

 We plot the column densities of six products of the irradiation as functions of the absorbed energy in Figure~\ref{Fig_prods}. We define the absorbed energy as $E_e \times F$, where $F = f \times t$ is the fluence measured in number of electrons per cm$^2$, and $t$ the irradiation time. Multiple Gaussian fitting has been used to compute \ce{C2O} and \ce{C3} column densities because of the blending with nearby features.

Increasing the electron energy, the maximum value of CO$_2$ column density $N^{(\rm m)}_{\ce{CO2}}$ increases, and shifts toward higher absorbed energies (see Figure \ref{Fig_prods}. For instance, in the 200 eV electron experiment, $N^{(\rm m)}_{\ce{CO2}} = 0.8$~ML is located at $6 \times 10^{16}$~eV~cm$^{-2}$, while in the 1000 eV experiment $N^{(\rm m)}_{\ce{CO2}} = 3.2$~ML is reached at $2 \times 10^{17}$~eV cm$^{-2}$. This behaviour is also observed for other species, see e.g., \ce{C3}, \ce{C5O} and \ce{C4O2}. 

\begin{figure}
\centering
\includegraphics[width=8cm]{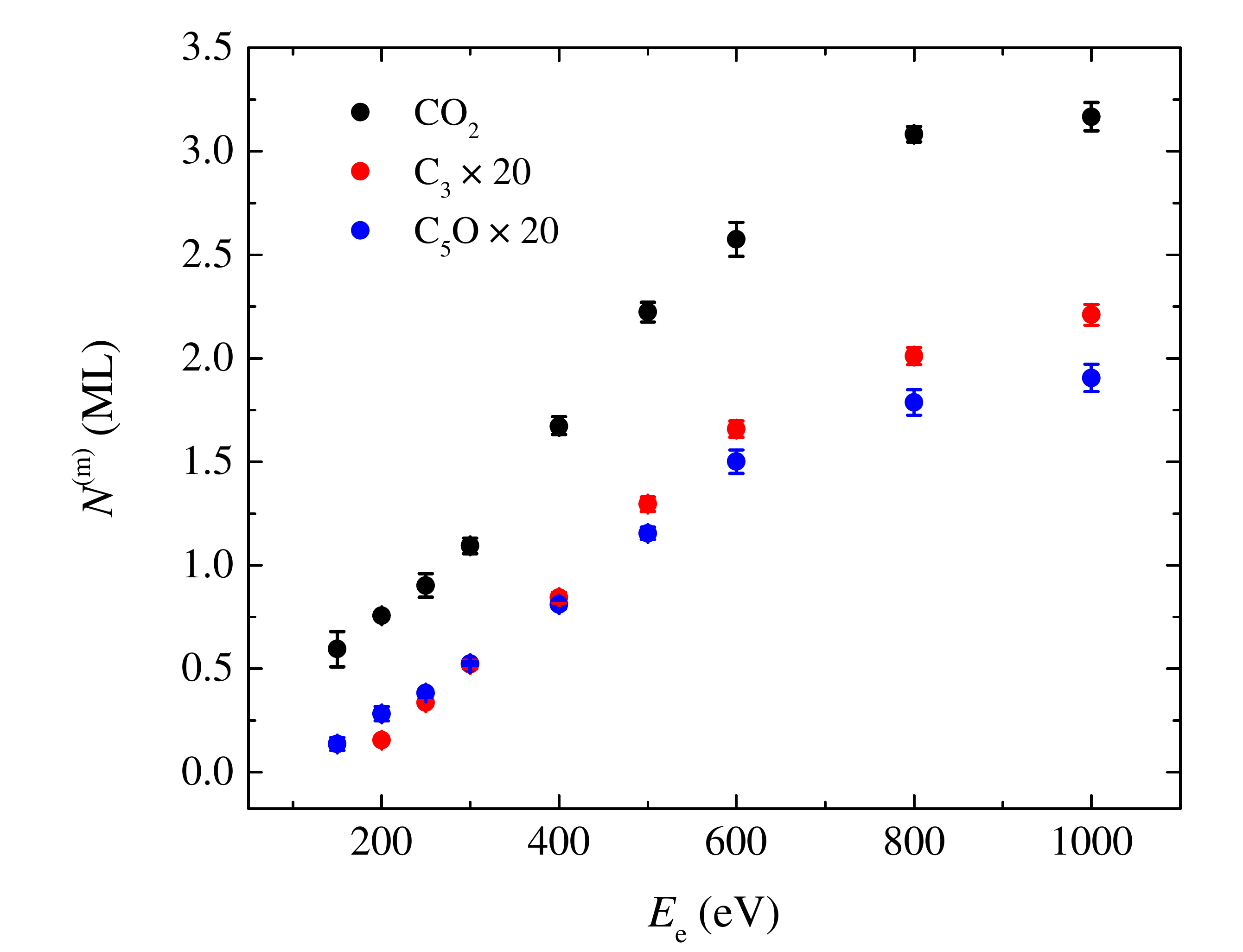}
\caption{Maximum column density of \ce{CO2}, \ce{C3}, and \ce{C5O} versus electron energy. The column densities for \ce{C3}, and \ce{C5O} are multiplied by factor of 20.} 
\label{Fig_CO2_max}
\end{figure}

The maximum values of the column densities as functions of the electron energy for \ce{CO2}, \ce{C3} and \ce{C5O} are shown in Figure~\ref{Fig_CO2_max}. The $N^{(\rm m)}$'s increase almost linearly with the electron energy up to 250~eV; then, at higher energies the linear regime breaks down and the maximum column densities tend to saturate.

\subsection{CO Destruction Cross-Section}
\begin{figure}
\centering
\includegraphics[width=8cm]{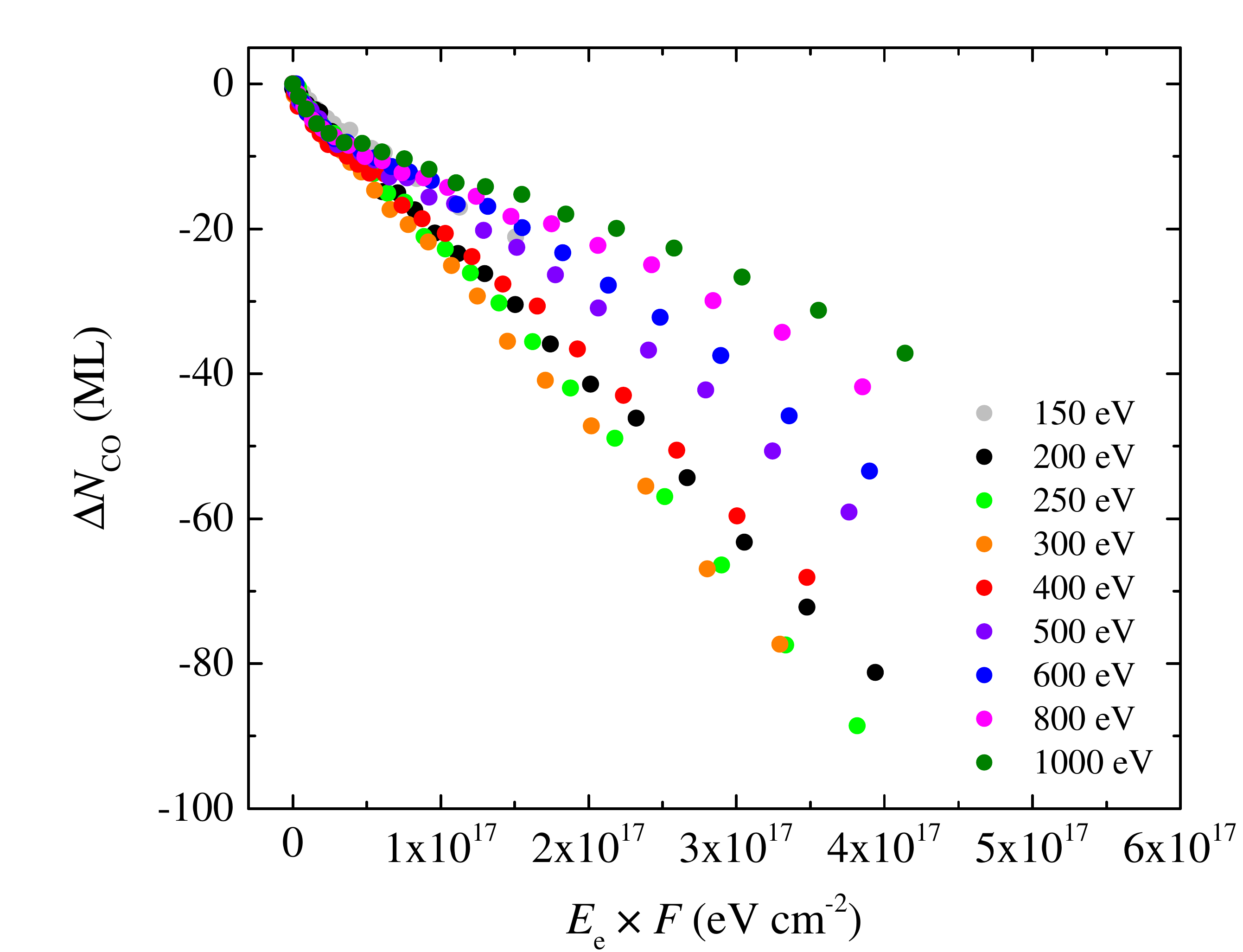}
\caption{CO column density estimated from $^{13}$CO versus the absorbed energy during the irradiation of 150 (grey dots), 200 (black), 250 (light green), 300 (orange), 400 (red), 500 (purple), 600 (blue), 800 (pink), and 1000 eV (dark green) electrons.}
\label{Fig_CO_destr}
\end{figure}

The destruction of CO during the irradiation is shown in Figure~\ref{Fig_CO_destr}. Since the CO stretching band at 2136 cm$^{-1}$ is close to saturation, we choose to compute CO column densities using the $^{13}$CO isotopologue.

The destruction cross-section of CO can be derived from the following equation
\begin{equation} \label{CO_Dest}
{\rm d}n_{\rm CO}(s,t) = - n_{\rm CO} \sigma_{D} f(s,t){\rm d}t
\end{equation}
where ${\rm d}n_{\rm CO}$ is the number density of destroyed CO molecules, $n_{\rm CO}$ the CO number density, $\sigma_{D}$ the destruction cross-section, and $f(s,t)$ the electron flux along the penetration path. Since only 4\% of initial number density of CO is converted into new species we assume $n_{\rm CO}$ constant within the ice. Integrating equation~(\ref{CO_Dest}) along the penetration path we obtain
\begin{equation} \label{Integ1}
{\rm d}N_{\rm CO} (t)  = -n_{\rm CO} \sigma_{D} ~ \int_{0}^{\Lambda_e} f(s,t)~{\rm d}s~{\rm d}t
\end{equation}
where we indicate with $N_{\rm CO}$ the CO column density. Finally the destruction cross-section is defined as
\begin{equation}
\sigma_{D} = -\frac{\Delta N_{\rm CO}}{n_{\rm CO}  \int_{0}^{\Lambda_e} F(s){\rm d}s}
\end{equation}
where $\Delta N_{\rm CO}$ is the column density of destroyed CO. We use CASINO to evaluate $\int_{0}^{\Lambda_e} F(s){\rm d}s$ for each electron energy. 

\begin{figure}
\scriptsize
\centering
\includegraphics[width=8cm]{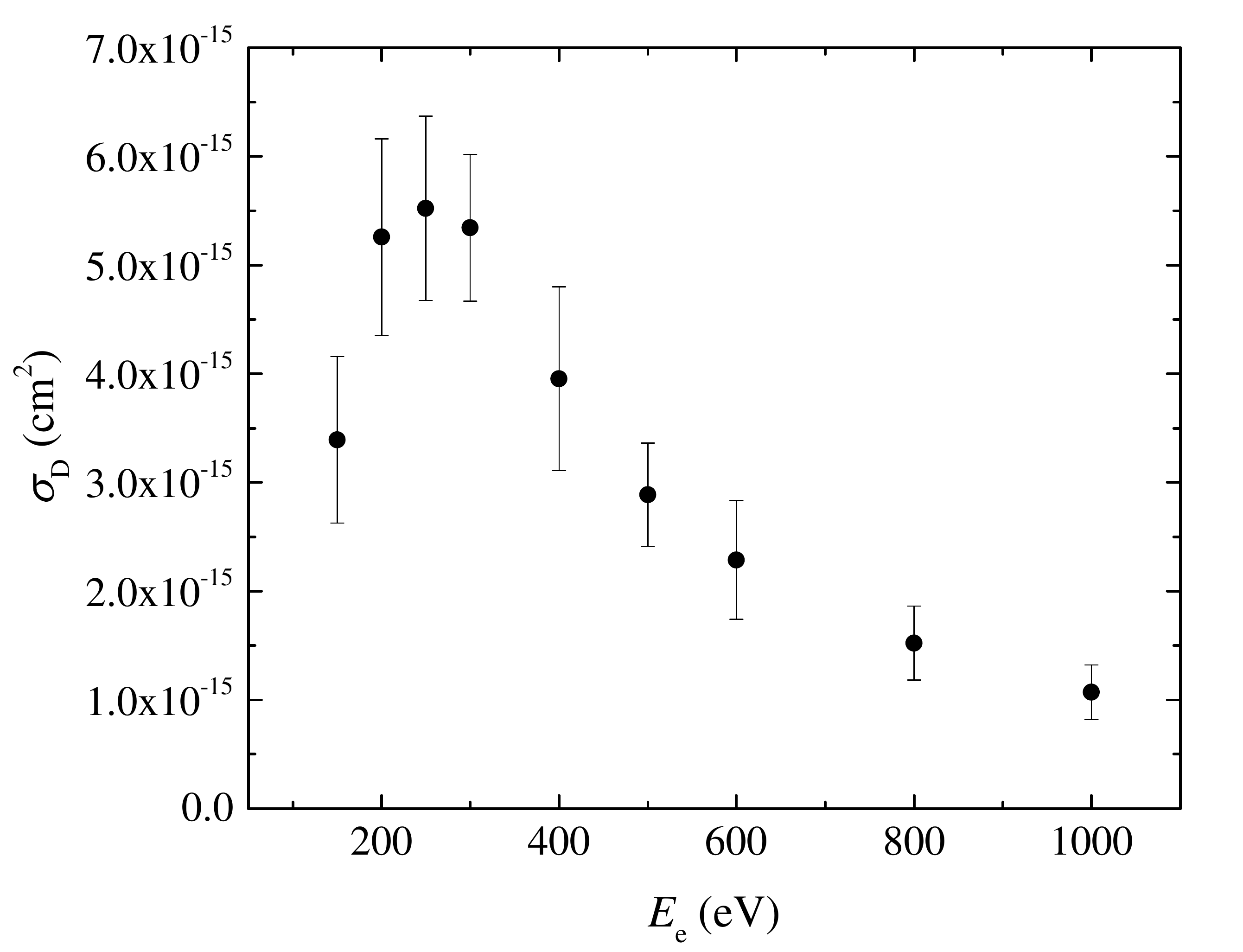}
\caption{CO destruction cross-section as a function of the electron energy. $\pm 1\sigma$ error bars are plotted.} \label{fig_CO_dest_y}
\end{figure}
The cross-section, $\sigma_{D}$, as a function of the electron energy is shown in Figure~\ref{fig_CO_dest_y}. This quantity has a broad peak centred around 250 eV, and it shows similarities in shape with the electron impact ionization cross-section of gas-phase CO (e.g., \citealt{Chu02}, and references therein), whose peak is instead located at a lower energy, 120~eV. The destruction cross-section for CO ice is about one order of magnitude larger than that obtained for gas-phase. 

\subsection{CO$_2$ Production Cross-Section}
\begin{figure}
\centering
\includegraphics[width=7.5cm]{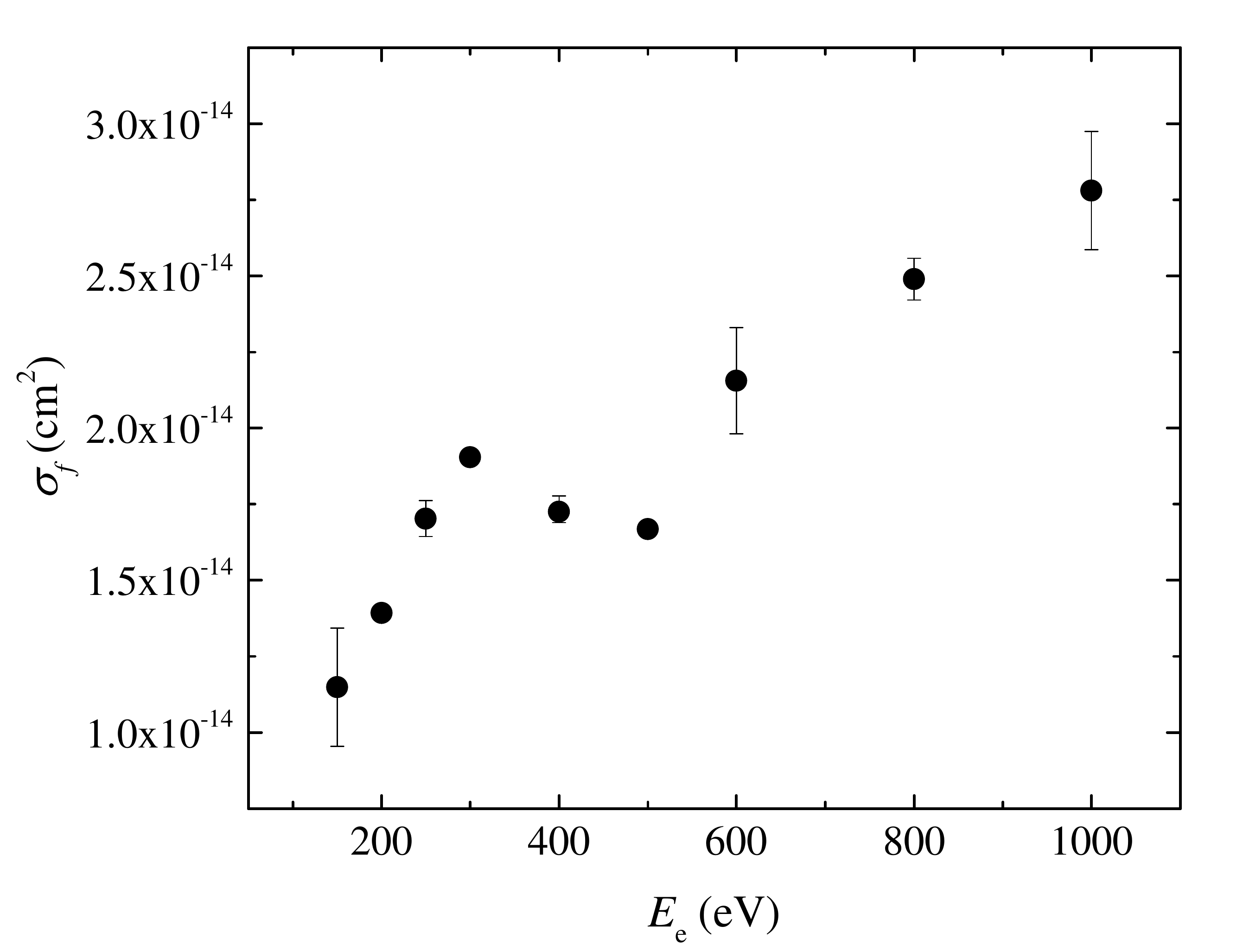}
\caption{\ce{CO2} production cross-section as function of the energy of the impacting electrons. $1~\sigma$ error bars are plotted.}
\label{Fig_CO2_PCS}
\end{figure}

We derive the production cross-section of carbon dioxide, $\sigma^{(1)}_f$ through a fitting relation that takes into account only production processes, i.e considering a first-order kinetic process \citep{Men10}
\begin{equation}
N_{\rm CO_2} = N^{(\rm m)}_{\rm CO_2} \times (1-e^{-\sigma^{(1)}_f F})
\label{Eq_form_CS}
\end{equation}
where $F$ is the number of electrons per cm$^2$ at a given irradiation time. This fit has been obtained using the column density profiles up to their maximum values $N^{(\rm m)}_{\rm CO_2}$. The resulting production cross-section is shown in Figure~\ref{Fig_CO2_PCS}. The cross-section exhibits a dip between 400 and 500 eV. Such feature has been confirmed by repeating the experiment, and fitting the results. We discuss this point in Section~\ref{disc}.

To test the robustness of this result, we also estimate the production cross-section using the consecutive reaction formalism
\begin{equation}
    2\,{\rm CO} \xrightarrow{\sigma_f} {\rm CO}_2 + {\rm C} \xrightarrow{\sigma_d} {\rm CO} + {\rm O} + {\rm C}
\end{equation}
\begin{equation}
N_{\rm CO_2} = N^\infty_{\rm CO_2} + A {\rm e}^{-\sigma^{(2)}_d F} - B {\rm e}^{-\sigma^{(2)}_f F} 
\label{due}
\end{equation}
where $N^\infty_{\rm CO_2}$ the asymptotic column density for very high fluences, $\sigma^{(2)}_f$ and $\sigma^{(2)}_d$ the production and destruction cross-sections,  and $A$ and $B$ two fitting parameters. We obtain values  very similar to $\sigma^{(1)}_f$. The destruction cross-section $\sigma^{(2)}_d = 2-3 \times 10^{-15}$~cm$^2$ should be taken as an order of magnitude estimate because of the limited energy range exploited to trace the decline of $N_{\ce{CO2}}$ (e.g., in 1000~eV experiment we have just 4 points after the peak).

\begin{deluxetable*}{ccccc}
\tabletypesize{\scriptsize} 
\tablecaption{Products\label{Assi}}
\tablewidth{45pt}
\tablehead{\\Wavenumber & Assignment & Band Strength & Lowest  Energy Detection & References\\
(cm$^{-1}$) & & (cm molecule$^{-1} \times 10^{-17}$) &(eV) &}
\startdata
3741    &\ce{C3O2}      &           &800       &\\
3707    &\ce{CO2}       &           &150        &\citet{G95,Ge96}\\
3601    &\ce{CO2}       &           &300        &\citet{G95,Ge96}\\
3069    &\ce{C3O2}      &           &150        &\citet{G01,MF64}\\
2399    &\ce{C3O2}      &0.8    &150        &\citet{G01,MF64}\\
2346    &\ce{CO2}       &7.6    &150        &\citet{Y64}\\
2330    &OC$^{18}$O     &           &150        &\citet{F87}\\
2281    &$^{13}$CO$_2$  &7.8    &150        &\citet{G95}\\
2248    &\ce{C3O}       &           &150        &\citet{DK71}\\
2243    &\ce{C3O2}      &1.3    &150        &\citet{MF64,HSL66},\\
        &               &           &           &\citet{G01}\\
2235    &---            &           &150        &\citet{Cia16}\\
2212    &\ce{C5O2}      &           &150        &\citet{H88,Mai88}\\
2194    &\ce{C3O2}      &           &150        &\citet{BR85}\\
2190    &\ce{C7O2} (\ce{C3O2})      &           &150     &\citet{Ma91}\\
2183    &\ce{C7O2}      &           &200        &\citet{Ma91}\\
2173    &---            &           &200        &\citet{Cia16}\\
2163    &\ce{C5} (\ce{C6O})         &           & 150     &\citet{DK71,Di76},\\
&               &           &           & \citet{Ma91}\\
2153    &---            &           &150        &\citet{Cia16}\\
2139$^\dagger$    &CO             &1.1    &  &\citet{Jia75}, \citet{Hu93}\\
2122    &\ce{C4O2} (\ce{C7O2})      &25.0    &150    &\citet{K98}, \citet{Mai88}\\
2115    &\ce{C7O2}      &           &150        &\citet{Ma91}\\
2106    &\ce{C5O}/\ce{C3O2}         &           &150    &\citet{Jam06}, \citet{G01}\\
2092$^\dagger$    &$^{13}$CO      &1.3    &  &\citet{G95}\\
2088$^\dagger$    &C$^{18}$O      &           &  &\\
2077    &\ce{C3O2} (\ce{C9}, \ce{C10})  &       &150    &\citet{Jam06, Van96}\\
        &                               &       &       &\citet{Fr97}\\
2065    &\ce{C8}        &           & 150        &\citet{Fr97}\\
2062    &\ce{C6O}/\ce{C7O} (\ce{C5O2})  &       &  150    &\citet{H88}, \citet{Mai88}\\
2058    &\ce{C5O2}      &           &  150        &\citet{Jam06,Mai88}\\
2049    &---            &           &  150        &\citet{Cia16}\\
2045    &---            &           &  150        &\citet{Cia16}\\
2019    &\ce{C3}        &10.0    &200        &\citet{H94}\\
1988    &\ce{C2O}       &           &  150        &\citet{Jam06}, \citet{DK71},\\
        &               &           &           &\citet{JM65}, \citet{Ge96}\\
1981    &\ce{C2O}       &           &  150        &\citet{Jam06}\\
1969    &\ce{C2O}       &           &  500           &\citet{JM65}\\
1951    &\ce{C6} (\ce{C11})         &           &250    &\citet{Kr92}, \citet{Tam97}\\
1939    &\ce{C10} (\ce{C11})        &           &  300    &\citet{Fr97}\\
1924    &\ce{C4O}       &           &200        &\citet{Mai88}\\
1918    &\ce{C4O}       &           &200     &\citet{Mai88}, \citet{Dib00}\\
1816    &\ce{C5O} (\ce{C12})        &6.2    &250    &\citet{Jam06}, \citet{D00}\\
1704	&\ce{C8}        &           &400        &\citet{Fr97}\\
1697	&\ce{C7O2}      &           &  400        &\citet{Ma91}\\
1592	&\ce{C9}        &           &  400        &\citet{Kr95}\\
1578	&---            &           &  500        &\citet{Cia16}\\
1563	&\ce{C3O2}      &           &400        &\citet{Wa02,Jam06} \\
1274	&\ce{CO2}       &           & 600        & \citet{Ga85} \\
1145	&\ce{C5O2}      &           &1000       &\citet{Mai88}\\
\enddata
\tablecomments{
$^\dagger$ parent molecule}
\end{deluxetable*} 

\subsection{CO and \ce{CO2} Desorption}
\begin{figure}
\centering
\includegraphics[width=8.5cm]{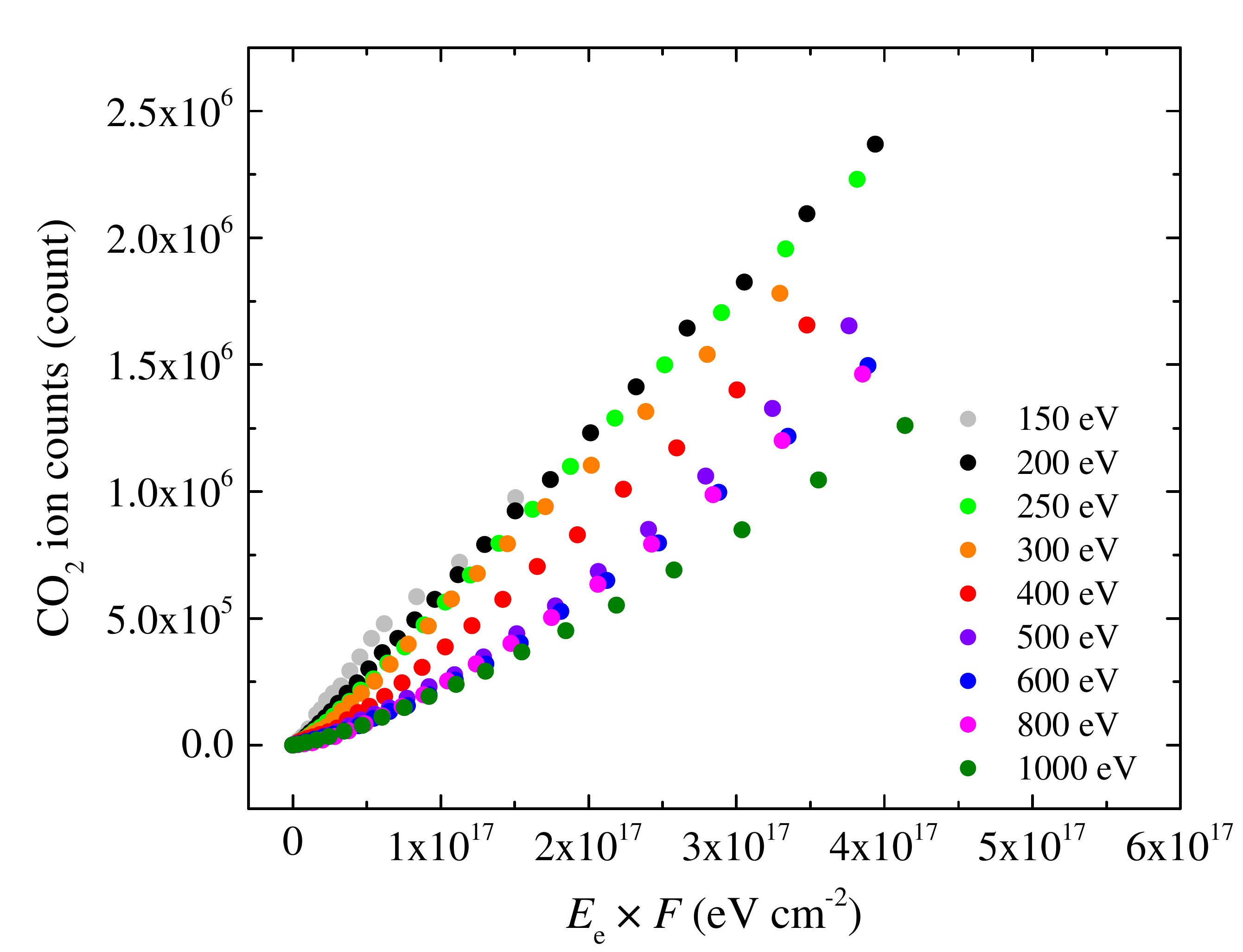}
\caption{\ce{CO2} ($m/z = 44$) accumulated ion counts as functions of the absorbed energy during radiolysis experiments exploiting electrons of 150 (grey dots), 200  (black), 250 (light green), 300 (orange), 400 (red), 500 (purple), 600 (blue), 800 (pink), and 1000 eV (dark green).}
\label{Fig_CO2_deso}
\end{figure}

\begin{figure*}
\centering
\begin{tabular}{cc}
\includegraphics[width=8.cm]{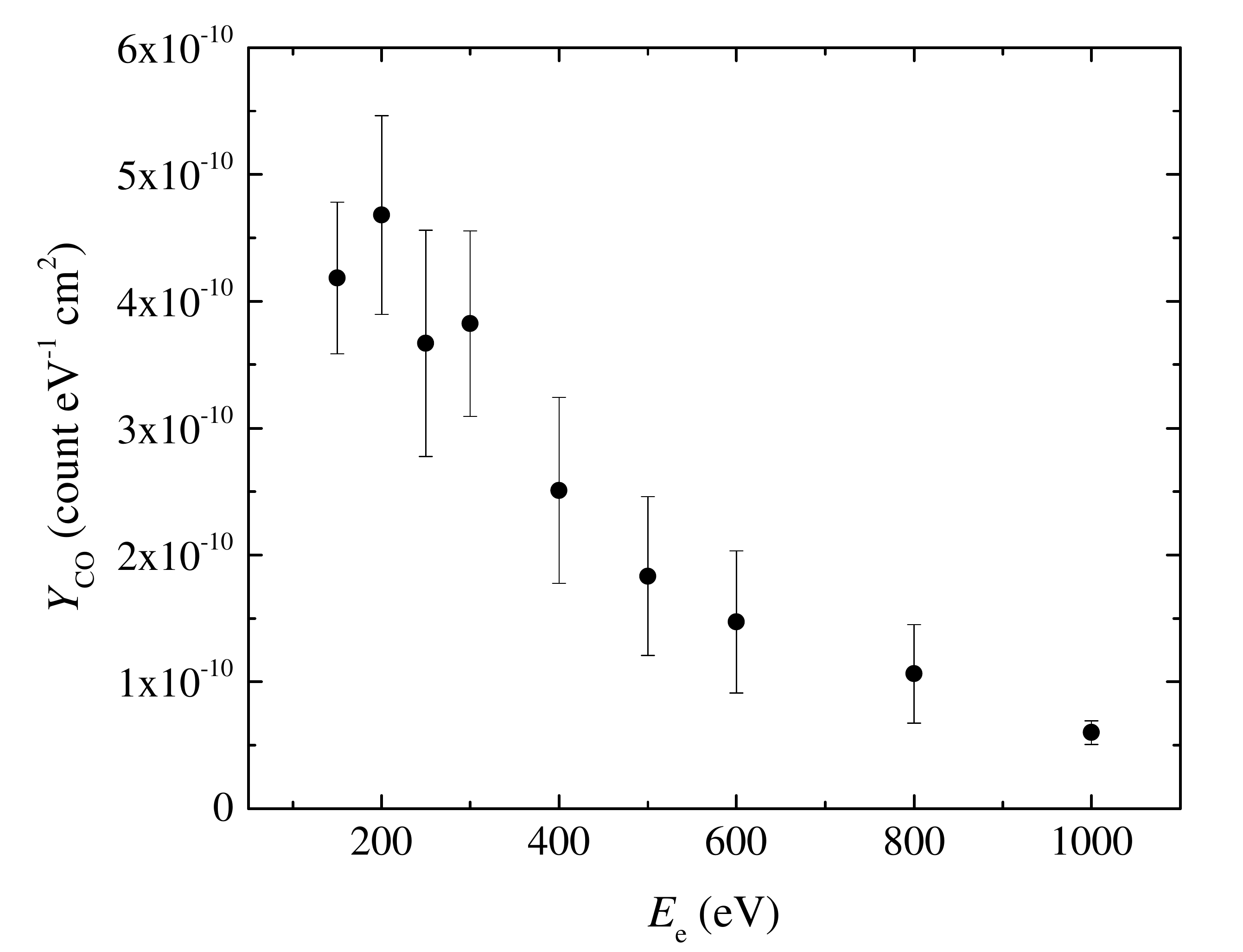} &
\includegraphics[width=8.cm]{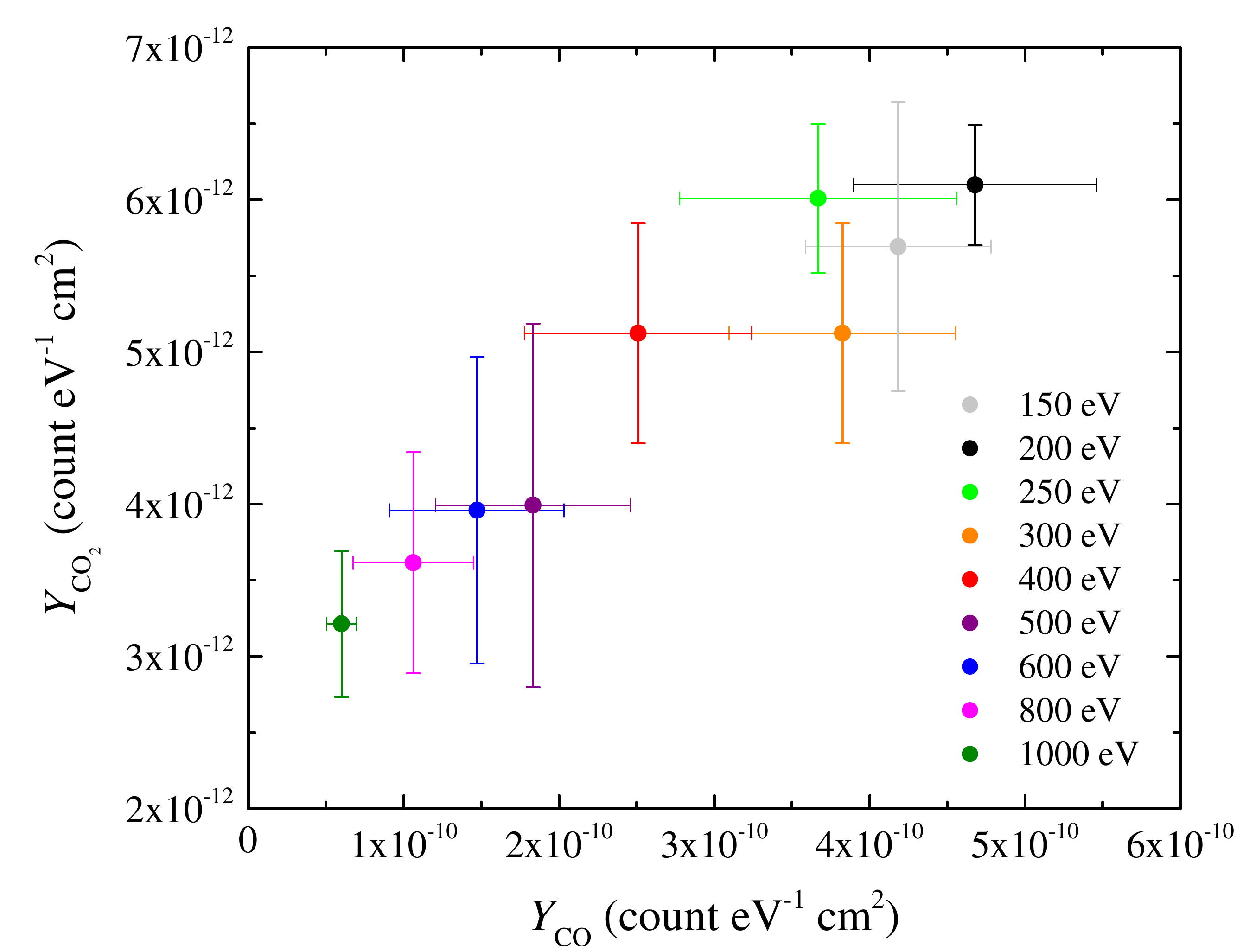}
\end{tabular}
\caption{Left panel: CO desorption yield as a function of the electron energy; right panel: comparison between CO and CO$_2$ desorption yields for 150 (grey), 200 (black), 250 (light green), 300 (orange), 400 (red), 500 (purple),  600 (blue), 800 (pink), 1000 (dark green) eV electrons. $\pm~1\sigma$ error bars are plotted.}
\label{COcomp2}
\end{figure*}

During radiolysis experiments the QMS detected a few species desorbing the ice, that include both the parent CO and products. Among the products, the strongest detected signal is coming from \ce{CO2} ($m/z = 44$), the major reaction product of the irradiation. We assume that this mass is produced entirely by \ce{CO2}. Such an approximation is justified by the negligible desorption of the second major product \ce{C3O2}. In Figure~\ref{Fig_CO2_deso} is shown the accumulated ion count profile for $m/z = 44$ as a function of the absorbed energy. The accumulated counts increase with increasing absorbed energy, and decrease with the energy of the impacting electrons, in qualitative agreement with the measurements of the electron-stimulated desorption yields performed by \citet{Tr07}.
We compute the average desorbing yield, $Y$ using the accumulated ion counts $\cal K$ as
\begin{equation}
Y (E_e) = \frac{1}{E_e} \bigg \langle \frac{{\rm d}{\cal K}}{{\rm d}F} \bigg \rangle_{\rm irr}
\label{Eq_CO_y2}
\end{equation}
In the left panel of Figure \ref{COcomp2} the desorption yield of CO decreases with increasing  electron energy.

The \ce{CO2} and CO desorption yields are compared in the right panel of Figure~\ref{COcomp2}. The yields are positively correlated suggesting that \ce{CO2} co-desorbs with the much more abundant CO.

\subsection{The Role of Ice Thickness}
\begin{figure}
\centering
\includegraphics[width=8cm]{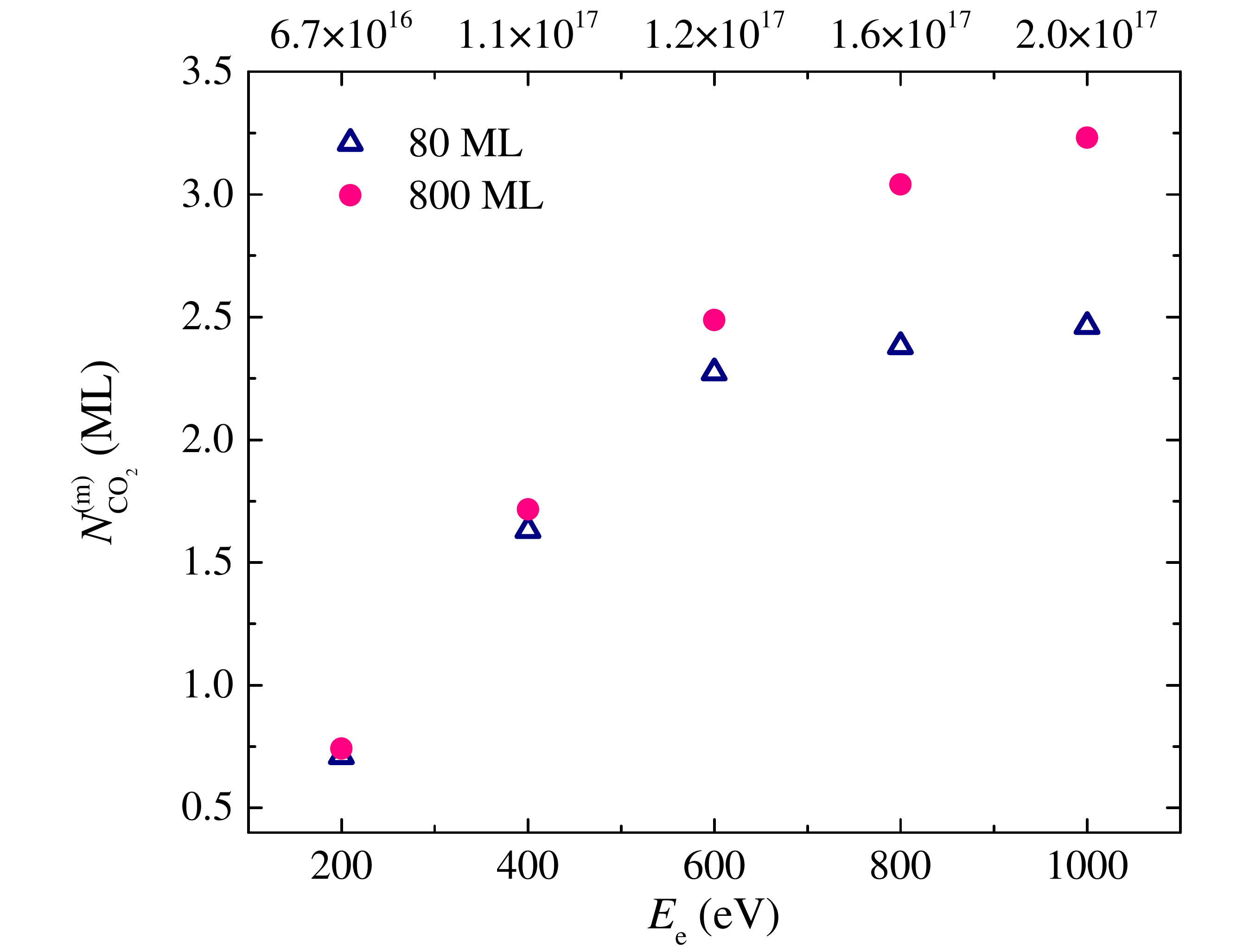}
\caption{Maximum \ce{CO2} column density as a function of the electron energy (bottom axis) and absorbed energy (top axis) for CO ice thicknesses of 80 (blue triangles) and 800 (red dots) ML.}
\label{Fig_CO2_100&1000}
\end{figure}

The effects of ice thickness on the chemical evolution has been explored by running additional experiments with electrons of 200, 400, 600, 800 and 1000 eV on a $76.9 \pm 0.5$ ML ($34.7 \pm 0.2$~nm) ice. The \ce{CO2} peak production is very similar for low energy electrons, while for energies higher than 600 eV, the production in thin ices declines. This is clearly proven in Figure~\ref{Fig_CO2_100&1000} where maximum \ce{CO2} column densities in both thin ($\sim 80$ ML) and thick ($\sim$ 800 ML) irradiated ices are compared. 

The difference in $N^{(\rm m)}_{\rm CO_2}$ increases with the energy of the electrons. For 1000~eV electrons $N^{(\rm m)}_{\rm CO_2}$ results are approximately 30\% higher in the thick irradiated ice. It is worth noting that for each electron energy, the maximum of \ce{CO2} column density occurs at the same absorbed energy for both thicknesses. A similar behaviour has been detected in 1~keV electron irradiation experiments of a pure \ce{NH3} ice \citep{Shul19}.

\begin{figure}
\centering
\includegraphics[width=8cm]{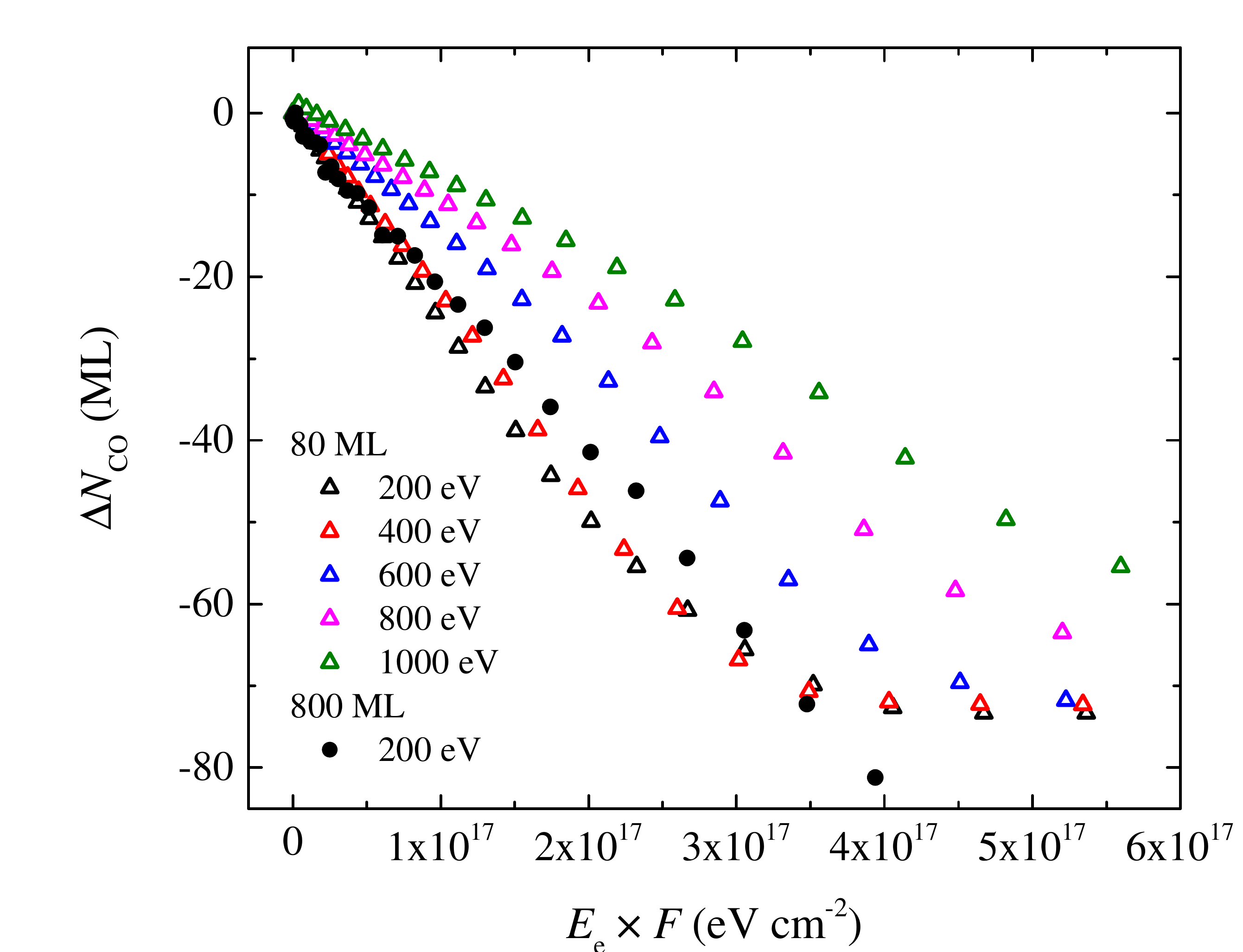}
\caption{Destroyed CO column density as function of absorbed energy for 200 (black), 400 (red), 600 (blue), 800 (pink), 1000 (dark green) eV electrons in 80 ML (empty triangle) ice. The solid black dots refer to 200 eV electrons in 800 ML ice as from Figure~\ref{Fig_prods}.}
\label{Fig_DCO_100&1000}
\end{figure}

The CO destruction in the thin ice experiments is reported in Figure~\ref{Fig_DCO_100&1000}. The behaviour of the CO column density as a function of the electron energy is similar to that observed in thick ices. Most destruction occurs in the lowest electron energy experiments. The main difference is the saturation at low energies where the ice is almost totally destroyed. As comparison we  report the 200 eV experiment in thick ice (black dots) in which at $3.5 \times 10^{17}$ eV cm$^{-2}$ only 10\% of the initial CO column density is destroyed and saturation does not occur. 

\section{Discussion} \label{disc}
In this work we study the chemical evolution of pure CO ices irradiated by electrons in the $150-1000$~eV range. The inventory of reaction products does not change with the energy of the impinging electrons. As for other energetic sources (e.g., ultraviolet photons or X-rays), the most abundant species is \ce{CO2}. The fraction of destroyed CO going into the formation of products increases with the electron energy. When \ce{CO2} reaches its peak abundance, this fraction is $0.17 \, (150~{\rm eV}) - 0.4 \, (1000~{\rm eV})$. For instance, in the 1000~eV experiment the \ce{CO2} peak column density is 3.2 ML against 20 ML of destroyed CO. This means that approximately 30\% of the destroyed CO is used for \ce{CO2}, while only 10\% for other (minor) products. The remaining 60\% is mostly desorbed, with a fraction dissociated into $\rm C + O$. According to \citet{Jam06} this fraction cannot be large, because O$_2$ and O$_3$, produced through barrierless reactions, would be otherwise detected (while they are not).

\ce{CO2} column densities exhibit a broad maximum, that raises and steepens with the energy of the impacting electrons. The absorbed energy (eV cm$^{-2}$) at which the \ce{CO2} column density peak is attained, increases as well.  

The results of the experiment performed with $E_e = 400$~eV appear to separate two different regimes. This behaviour is reflected by the shape of the \ce{CO2} production cross-section (see Figure \ref{Fig_CO2_PCS}) that shows a peak at 300~eV, a dip at about 500 eV, and a convex rise towards higher energies. The peak energy is close to the maximum of the CO destruction cross-section (Figure \ref{fig_CO_dest_y}). The cross-section after reaching its maximum value at about 250~eV, decreases steadily for higher energies. This might appear inconsistent with the high energy rise of \ce{CO2} production cross-section. These two seemingly conflicting results may be reconciled once we take into account CO and \ce{CO2} desorption induced by the electron irradiation. 

\begin{figure}
\centering
\includegraphics[width=8cm]{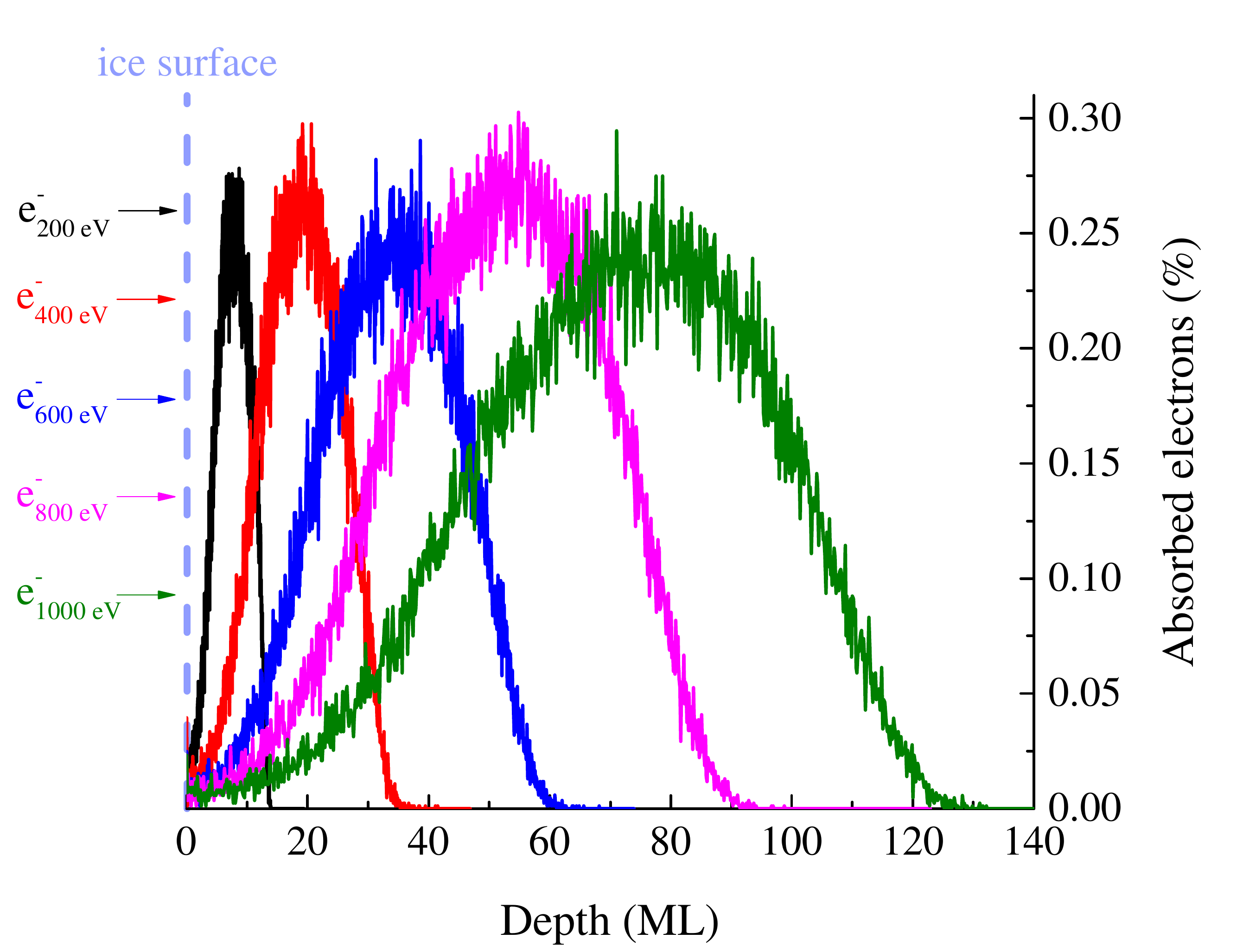}
\caption{The histogram of electron distributions as obtained from CASINO across the CO ice. The right vertical axis shows the percentage of the absorbed electrons. Different colors indicate different electron energies, see left vertical axis. The number of bins in each histogram is 1000, and the total number of electrons is normalised to unity.}
\label{Fig_casino}
\end{figure}

In Table~\ref{exp} we list the electron penetration depth, $\Lambda_e$ for each electron energy. This quantity, derived using the CASINO software, increases quadratically with the energy of the impacting electron, and it is a measure of the ice thickness that is being processed by electrons of specific energies. The absorbed electron distribution within the ice has been computed from CASINO and it is shown in Figure~\ref{Fig_casino}. As the figure shows, the electron distribution inside the ice displays an asymmetric bell-shaped profile, peaking at a distance from the ice edge $\sim 2/3 \times \Lambda_e$. The \ce{CO2} produced during the irradiation follows the electron distribution. Thus, the \ce{CO2} production is increasingly affected by desorption, as the energy of the impacting electrons decreases.

\begin{figure}
\centering
\includegraphics[width=8cm]{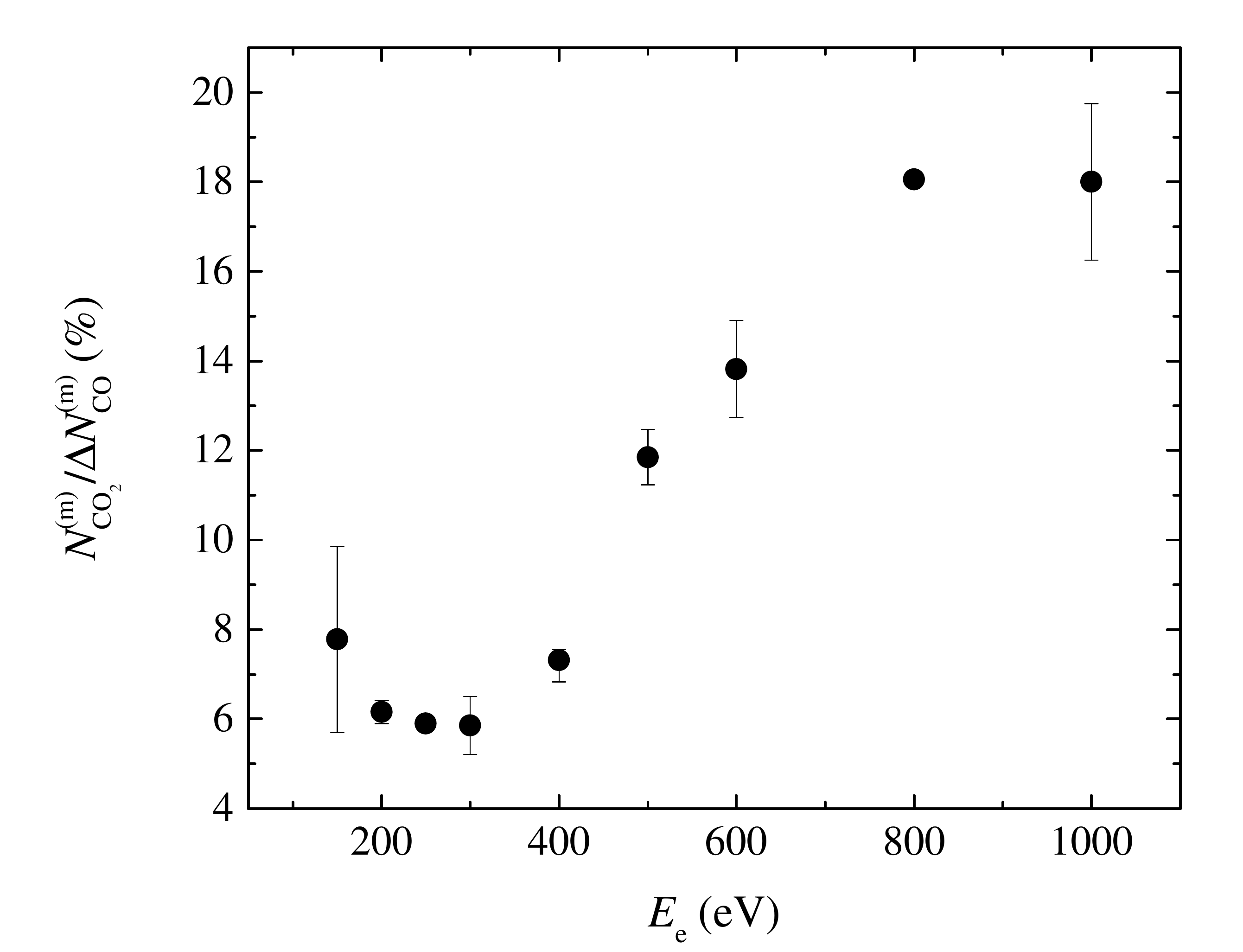}
\caption{The ratio between the maximum column density of produced \ce{CO2} ($N^{\rm (m)}_{\rm CO_2}$) and the corresponding destroyed column density of CO ($\Delta N^{(m)}_{CO}$) as a function of the electron energy $E_e$.}
\label{Fig_CO2_DCO_ratio}
\end{figure}

At a specific location in the ice, the \ce{CO2} concentration is determined by both chemistry and ice erosion. If the corresponding timescales are comparable, the measured \ce{CO2} column density is continuously produced over the desorption timescale, as new ice layers are processed by electrons. In fact, at low electron energies where the timescale of production and erosion are comparable we detect shallow peaks followed by a smooth decline of \ce{CO2} (see Figure~\ref{Fig_prods}). As a consequence, the \ce{CO2} production cross-section mirrors the trend in CO destruction, i.e. peaking at about 300~eV and then declining. In this regime the maximum produced \ce{CO2} column density, $N^{\rm (m)}_{\rm CO_2}$ is directly proportional to the column density of destroyed CO, $\Delta N^{\rm (m)}_{\rm CO}$. As the electron energy increases, the electron distribution widens (see Figure~\ref{Fig_casino}), decreasing significantly the number of electrons close to the ice edge, reducing desorption. The average power density $P_e = E_e \times f/\Lambda_e$ decreases due to the much larger spread along the electron penetration path. As a consequence, the \ce{CO2} column density begins to accumulate, following the increase of ice layers involved in the \ce{CO2} production with respect to the ones removed by desorption events. Thus, the production cross-section rises with the electron energy in response to the increase of $\Lambda_e$, and so does the $N^{\rm (m)}_{\rm CO_2}/\Delta N^{\rm (m)}_{\rm CO}$ ratio (Figure~\ref{Fig_CO2_DCO_ratio}).

\begin{figure}
    \centering
    \includegraphics[width=8cm]{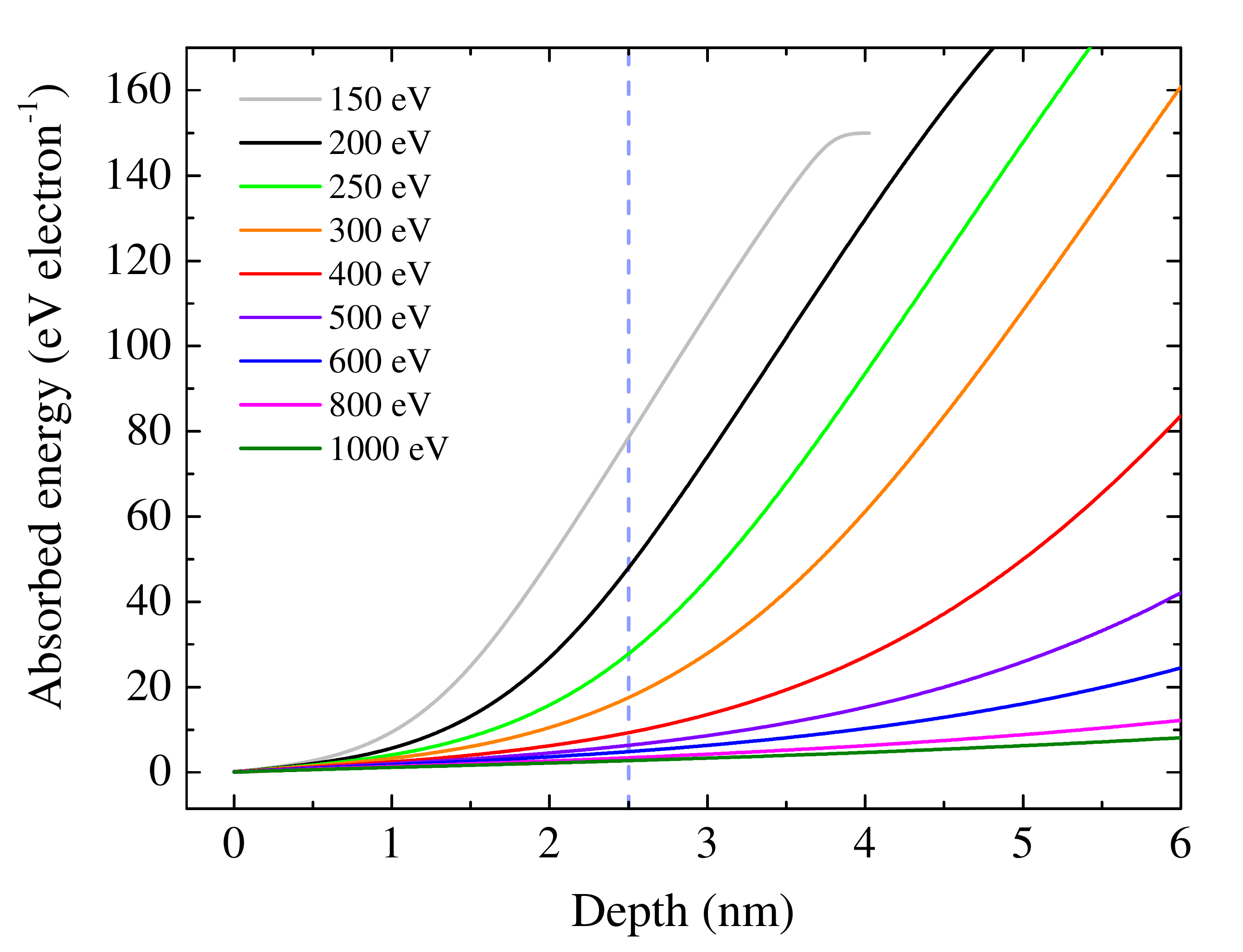}
    \caption{Energy deposited by electrons of different energies close to the ice boundary.}
    \label{Fig_ED25}
\end{figure}

In Figure \ref{Fig_ED25} we show the energy deposited by electrons of different energies as a function of the penetration within the ice. For all locations the amount of deposited energy decreases monotonically with increasing energy of the impacting electrons. At 2.5~nm, a distance lower than the penetration depth of the lowest energy electron considered in this work, 150~eV electrons deposit 78 eV, while 1000~eV electrons just 3~eV.

Finally, when the thickness of the ice is comparable with the penetration depth, the abundances of products decrease with respect to the case of a thicker ice. 

For electrons with higher energies, the penetration depth increases significantly. In the case of 5~keV electrons used by \citet{Jam06} the penetration depth, $\Lambda_e \sim 1500$~ML, is about one order of magnitude larger than for 1~keV electrons (see Table~\ref{exp}). Extrapolating the results of our experiments to the case of 5~keV electron irradiation we may expect a significantly lower desorption (per eV), consequently a richer chemical production, and the \ce{CO2} column density peak located at a much larger absorbed energy. This is qualitatively in agreement with the findings of \citet{Jam06}. 

We compare our results with those found by \citet{Jam06}. The two experiments differ in the sample sizes that are 3~cm$^2$ and ($2\pm0.1$~cm$^2$), in \citet{Jam06} and in the present work, respectively. Moreover, \citet{Jam06} scanned over the sample area while we irradiate in a single spot. Extrapolating the data shown in Figure~\ref{Fig_CO2_100&1000} up to 5~keV, we obtain $N^{(\rm m)}_{\rm CO_2} \sim 5.5$~ML. \citet{Jam06} derive a best-fit value for this quantity as $N^{(\rm m)}_{\rm CO_2} \sim 5.8$~ML, in remarkable agreement with the present results. 
 
\section{Conclusions}\label{concl}
In this work we have performed electron impact experiments on pure CO ices. Both, CO ices and fast electrons are ubiquitous in space, from dense molecular clouds to planetary surfaces. We employ electrons in the middle-energy range ($150-1000$~eV) in order to fill the gap existing in previous experiments adopting relatively high energy electrons \citep{Jam06} or studying low-energy ($1-30$~eV) electron interactions with CO such as dissociative electron attachment \citep{M12}. Our assumed range of energy is also close to the energy distribution of primary electrons emitted in the interaction of cosmic-rays and X-rays with molecular gas \citep{D99}.  

The main results of this work are as follows:
\begin{enumerate}
    \item the identification of new formed species provides an inventory very close to the one resulting from X-ray irradiation, confirming the driving role of secondary electron cascade in the chemistry of the ice;
    \item we provide the evolutionary tracks of products and find that major products (i.e. \ce{CO2}, \ce{C3}, \ce{C5O}) evolve following similar patterns;
    \item CO desorption is relevant, eroding significantly the ice upon long irradiation time; 
    \item CO destruction cross-section, \ce{CO2} production cross-section, CO and \ce{CO2} desorption yields have been determined; the shape of \ce{CO2} production cross-section reflects the competition between chemistry and desorption;
    \item \ce{CO2} molecules (and presumably all the other products) co-desorb with CO ice, with the yields declining with increasing electron energy (penetration depth); 
    \item extrapolation of the present results to higher energy provide a good agreement with the results of 5~keV electron experiments performed by \citet{Jam06}.
\end{enumerate}

The chemistry that we observe to occur in CO ice under electron processing may have implication in the atmospheric photochemistry of cold planets hosting surface CO ices. A few years ago, \citet{L10} discovered a seasonal evolution of Triton's atmosphere over decades. Something similar has been observed taking place on Pluto \citep{BF16}. Volatile chemicals such as nitrogen, methane and carbon monoxide start out as ices on the surface, then sublime into the atmosphere when temperatures rise. Solar wind electrons  \citep{Hoog96} and/or primary electrons produced by cosmic-rays inducing a chemistry such as the one we have described in this work, may contribute to enrich the nitrogen-rich atmosphere of these small planets with suboxides and carbon chains (in addition to CO$_2$). As volatile gases drift upwards, photochemical reactions create new carbon and nitrogen compounds. Moreover, carbon chains at the boundary layer may assemble and combine into solid carbonaceous matter at cryogenic temperature, as has been shown in a recent experiment by \citet{Fulvio17}, forming nano-sized particles. These particles settle down onto the planets's surface, and coat it until warming begins and they are lofted into the atmosphere.

\acknowledgments
This work has been supported by the project MOST grants MOST 107-2112-M-008-016-MY3 (Y.-J.C.), Taiwan. We also acknowledge support from INAF through the PRIN-INAF 2016 The Cradle of Life - GENESIS-SKA (General Conditions in Early Planetary Systems for the rise of life with SKA).

\end{document}